# Connection between coherent states and some integrals and integral representations

**Dušan POPOV**

University Politehnica Timisoara, Romania,
Department of Physical Foundations of Engineering
Bd. V. Pârvan No. 2, Timișoara 300223
dusan_popov@yahoo.co.uk
ORCID: 0000-0003-3631-3247

## Abstract

The paper presents an interesting mathematical feedback between the formalism of coherent states and the field of integrals and integral representations involving special functions.. This materializes through an easy and fast method to calculate integrals or integral representations of different functions, expressible by means of Meijer's G-, as well as hypergeometric generalized functions. The feedback starts from a fundamental integral that comes from the decomposition of the unity operator in the language of coherent states from quantum mechanics. In this way, integrals and integral representations are obtained, some that do not appear in the literature, and others already known, which can be verified by orthodox methods. All calculations are made using the properties of the diagonal operators ordering technique (DOOT), a relatively new technique of normal ordering of the creation and annihilation operators in quantum mechanics. The paper contributes to increasing the number of solvable integrals involving special functions.

## 1. Introduction

As a concept, the formalism of coherent states (CSs) was introduced almost a century years ago from Schrödinger [1], but only after a few decades did it return to the attention of researchers, finding numerous applications in various fields of physics (in quantum mechanics, quantum optics, quantum information theory and practice). Initially the concept of CSs was referred only to the one-dimensional quantum harmonic oscillator (HO-1D), and later it was extended to other quantum systems (anharmonic oscillators or systems). Today there is a wide variety of CSs (see, e.g. [2], [3], [4], [5], [6], [7], [8], [9]), which differ in their way of definition, they can still be classified into three large categories: Barut-Girardello (BG-CSs) [2], Klauder-Perelomov (KP-CSs) [3] and Gazeau-Klauder (GK-CSs) [4]. For the HO-1D all three manners of defining the CSs lead to the same results, i.e. to the same expression of CSs, and they are also named the *canonical* or *linear* coherent states. But for other quantum systems the three manners of building CSs lead to different expressions and these CSs are named *nonlinear*.

In the paper we present some primary elements, necessary to define CSs, using elements of the diagonal operator ordering technique (DOOT). Special attention is paid to the relationship of solving the unit operator, i.e. to its decomposition with respect to the projectors of the coherent states. According to the DOOT rules, inside the signs # #, the normally ordered operators can be seen as simple numbers, and so we can obtain several general integrals and integral representations implying generalized hypergeometric functions. Their verification is done by particularizing the indices and constants of these functions, which leads to new, as well as known results, obtained by other methods. Implicitly, the results show that the formalism of coherent states, corroborated with the DOOT technique can be useful also for deducing new integrals or integral representations from complex or real space.

## 2. Preliminaries on the coherent states formalism

Let us present, briefly, the main notions related to CSs.

Individually, an arbitrary set of CSs $|z>$ is labeled by a complex number $z=|z|\exp(i\varphi)$, $|z|\leq R\leq\infty$, $\varphi\in(0,2\pi)$, and can be developed in the Fock-vectors basis $|n>$ (having finite or infinite size, $M\leq\infty$). However, the most general expression for CSs is the so-called *generalized coherent states*, defined so that their normalizing function is a generalized hypergeometric function (G-HGF) ${}_pF_q\left(\{a_i\}_1^p\ ;\ \{b_j\}_1^q;|z|^2\right)$ [5]:

$$|z>=\frac{1}{\sqrt{{}_pF_q\left(\{a_i\}_1^p\ ;\ \{b_j\}_1^q;|z|^2\right)}}\sum_{n=0}^{M\leq\infty}\frac{z^n}{\sqrt{\rho_{p,q}(b/a\,|\,n)}}|n> \tag{2.1}$$

The generalized hypergeometric function ${}_pF_q\left(\{a_i\}_1^p\ ;\ \{b_j\}_1^q;|z|^2\right)$ is defined as

$${}_pF_q\left(\{a_i\}_1^p\ ;\ \{b_j\}_1^q;|z|^2\right)=\sum_{n=0}^{\infty}\frac{\prod_{i=1}^{p}(a_i)_n}{\prod_{j=1}^{q}(b_j)_n}\frac{\left(|z|^2\right)^n}{n\,!} \tag{2.2}$$

Here $\rho_{p,q}(b/a\,|\,n)$ is so called *structural constants* (because it determines the internal structure of CSs) which depends on the positive integers $p$ and $q$, as well as the sets of real or complex numbers $a\equiv\{a_1,a_2,...,a_p,\}\equiv\{a_i\}_1^p$ and $b\equiv\{b_1,b_2,...,b_q,\}\equiv\{b_j\}_1^q$.

Furthermore, all CSs $|z>$ must satisfy some conditions, named Klauder's minimal prescriptions [4]: a) *the continuity of labeling*, i.e. if $z'\to z$, then $\|\,|z'>-|z>\|\to 0$; b) the *normalization* but *non orthogonality*:

$$< z\,|\,z'> = \frac{{}_pF_q\left(\{a_i\}_1^p\ ;\ \{b_j\}_1^q\,; z^* z'\right)}{\sqrt{{}_pF_q\left(\{a_i\}_1^p\ ;\ \{b_j\}_1^q\,;|\,z\,|^2\right)}\sqrt{{}_pF_q\left(\{a_i\}_1^p\ ;\ \{b_j\}_1^q\,;|\,z'|^2\right)}} = \begin{cases} 1, & \text{for } z' = z \\ \neq 0, & \text{for } z' \neq z \end{cases} \tag{2.3}$$

and c) the most important condition, called *the completion relation* or *the resolution of the unity operator* into the CSs projectors

$$\int d\mu(z)|\,z >< z\,| = 1 \tag{2.4}$$

with the corresponding integration measure

$$d\mu(z) = \frac{d^2 z}{\pi}\,h(|\,z\,|) = \frac{d\varphi}{2\pi}\,d(|\,z\,|^2)\,h(|\,z\,|) \tag{2.5}$$

having a positive weight function $h(|\,z\,|)$. This must be determined separately, for each examined quantum system and each kind of CSs.

Let's also point out that the generalized hypergeometric functions ${}_pF_q\left(...\,;\,...;|\,z\,|^2\right)$ can be expressed through the Meijer's *G*-functions, $G_{p,q}^{m,n}\left(|\,z\,|^2\,|...\right)$ as [9]:

$${}_pF_q\left(\{a_i\}_1^p\,;\,\{b_j\}_1^q\,; A|\,z\,|^2\right) = \frac{\prod_{i=1}^{p}\Gamma(a_i)}{\prod_{j=1}^{q}\Gamma(b_j)}\,G_{p,q+1}^{1,p}\left(-A|\,z\,|^2\,\middle|\,\begin{matrix}\{1-a_i\}_1^p; & / \\ 0\quad ; & \{1-b_j\}_1^q\end{matrix}\right) \tag{2.6}$$

Now, let as consider a quantum system with a dimensionless Hamiltonian $\hat{H}$, having only the discrete energy spectrum, with ascendant and non-degenerate eigenvalues $e(n)$, i.e. $e(0) = 1 < e(1) < e(2) < ... < e(n)$, and the eigenvalue equation of Fock vectors $|\,n>$:

$$\hat{H}\,|\,n> = e(n)\,|\,n> \tag{2.7}$$

Particularly, we choose a pair of two Hermitical operators, in a particular way, the creation $\hat{E}_+$, and the annihilation $\hat{E}_-$, which acts on the Fock vectors $|\,n>$ so that their normal product is exactly equal to the Hamilton operator of the system:

$$\begin{matrix}\hat{E}_+\,|\,n> = \sqrt{e\,(n+1)}\,|\,n+1> \\ \hat{E}_-\,|\,n> = \sqrt{e\,(n)}\,|\,n-1>\end{matrix}\quad ,\quad \hat{E}_+\hat{E}_-\,|\,n> = e(n)\,|\,n>\quad ,\quad \hat{H} = \hat{E}_+\hat{E}_- \tag{2.8}$$

We suppose that the dimensionless energy eigenvalues have the following general expression (which will be implicitly motivated in the below) [10]:

$$e(n) = n\frac{\prod_{j=1}^{q}\left(b_j - 1 + n\right)}{\prod_{i=1}^{p}\left(a_i - 1 + n\right)} \quad , \quad n = 1, 2, 3, ... , \tag{2.9}$$

In this situation, the structural constants becomes

$$\rho_{p,q}(b/a\,|\,n) \equiv \prod_{l=1}^{n} e(l) = n!\frac{\prod_{j=1}^{q}\left(b_j\right)_n}{\prod_{i=1}^{p}\left(a_i\right)_n} \tag{2.10}$$

where $(x)_n = \Gamma(x+n)/\Gamma(x)$ is the Pochhammer symbol and $\Gamma(x)$ - the Euler's Gamma function.

The repeated action of the creation operator $\hat{E}_+$ on the vacuum (or ground) state $|\,0>$ leads to the following matrix relation

$$\begin{pmatrix} |\,n> \\ <n\,| \end{pmatrix} = \frac{1}{\sqrt{\rho_{p,q}(b/a\,|\,n)}} \begin{pmatrix} \left(\hat{E}_+\right)^n |\,0> \\ <0\,|\left(\hat{E}_+\right)^n \end{pmatrix} \tag{2.11}$$

A few years ago, Fan et all [11] have introduced a new calculation technique for normally ordered canonical creation and annihilation operators $\hat{a}^+$ and $\hat{a}$, called the *integration with ordered products* (IWOP), applicable only to one dimensional harmonic oscillator (HO-1D), which leads to some new mathematical results. Later, we have generalized the IWOP technique and applied it to the arbitrary quantum pair of operators $\hat{E}_+$, and the annihilation $\hat{E}_-$. We named this technique the *diagonal operators ordering technique* (DOOT), and we obtained also a series of useful results [10], [12], [13].

In short, the working rules of DOOT technique (using the sign # #) can be summarized as follows: Inside the DOOT sign # # the operators $\hat{E}_+$ and $\hat{E}_-$ *commute*, forming a product with normal ordering (with $\hat{E}_+$ on the left and $\hat{E}_-$ on the right), i.e. $\#\hat{E}_+\hat{E}_-\#$ (or its natural powers # $\left(\hat{E}_+\hat{E}_-\right)^m$ #) is formed. Inside the sign # #, these operators are considered to be *c-numbers*, so that algebraic operations can be performed according to the usual rules [10]. An additional consequence of using the DOOT is that they can be obtained a series of new integration relations, as well as integral representations, involving Meijer's $G_{p,q}^{m,n}\left(|\,z\,|^2\,|...\right)$ and generalized hypergeometric functions ${}_pF_q(... \,; ...;|\,z\,|^2)$ [14].

The completion relation (or the identity operator decomposition) for Fock - vectors, $\sum_{n=0}^{M\leq\infty} |\,n><n\,| = 1$, leads to the derivation of the expression of the projector on the vacuum state [10]:

$$|0><0|=\frac{1}{\#_pF_q\left(\{a_i\}_1^p\ ;\ \{b_j\}_1^q\ ;\ \hat{E}_+\hat{E}_-\right)\#} \tag{2.12}$$

which is *the same for all three kinds of CSs*, i.e. it is independent of the definition of CSs.

For a certain quantum system it can be defined three kinds of CSs: a) the Barut-Girardello CSs (BG-CSs); b) the Klauder-Perelomov CSs (KP-CSs), and c) the Gazeau-Klauder CSs (GK-CSs), whose expressions are different or divergent, excepting the case of HO-1D where these three definitions are convergent, i.e. their results are identical.

To find the integration measure, we have to solve an integral moment problem (the so called Stieltjes or Haussdorff moment, depending if $R=\infty$ or $R<\infty$). The convergence radius $R$ for any kind of CSs is determined by calculating the limits using the power series convergence criteria, for example:

$$R=\frac{1}{\lim\limits_{n\to\infty}\sqrt[n]{\rho_{p,q}(b/a\,|\,n)}} \quad ; \quad R=\lim_{n\to\infty}\frac{\rho_{p,q}(b/a\,|\,n)}{\rho_{p,q}(b/a\,|\,n+1)} \tag{2.13}$$

and the CSs exists (that is, they have a physical sense) only if $R\neq 0$.

The most important characteristics of the three types of CSs are the following:

***a) The Barut-Girardello coherent states (BG-CSs)*** are the eigenfunctions of the lowering operator $\hat{E}_-$ [2]:

$$\hat{E}_-\,|\,z>_{BG}=z|\,z>_{BG} \tag{2.14}$$

Their expansion in the Fock-vectors basis is

$$|\,z>_{BG}=\frac{1}{\sqrt{{}_pF_q\left(\{a_i\}_1^p\ ;\ \{b_j\}_1^q\ ;|\,z\,|^2\right)}}\sum_{n=0}^{\infty}\frac{z^n}{\sqrt{\rho_{p,q}(b/a\,|\,n)}}|\,n>\ . \tag{2.15}$$

The integration measure is obtained using the general relation for the classical integral from one Meijer's G-function [9]

$$\int_0^\infty dx\,x^{s-1}\,G_{p,q}^{m,n}\left(\omega x\left|\begin{array}{ll}\{a_i\}_1^n\ ; & \{a_i\}_{n+1}^p\\ \{b_j\}_1^m\ ; & \{b_j\}_{m+1}^q\end{array}\right.\right)=\frac{1}{\omega^s}\frac{\prod\limits_{j=1}^{m}\Gamma(b_j+s)\prod\limits_{i=1}^{n}\Gamma(1-a_i-s)}{\prod\limits_{j=m+1}^{q}\Gamma(1-b_j-s)\prod\limits_{i=n+1}^{p}\Gamma(a_i+s)} \tag{2.16}$$

The integration measure for BG-CSs is

$$d\mu_{BG}(z)=\frac{\prod\limits_{i=1}^{p}\Gamma(a_i)}{\prod\limits_{j=1}^{q}\Gamma(b_j)}\frac{d\varphi}{2\pi}d\left(|\,z\,|^2\right){}_pF_q\left(\{a_i\}_1^p\ ;\ \{b_j\}_1^q\ ;|\,z\,|^2\right)G_{p,q+1}^{q+1,0}\left(|\,z\,|^2\left|\begin{array}{ll}/\ ; & \{a_i-1\}_1^p\\ 0,\ \{b_j-1\}_1^q\ ; & /\end{array}\right.\right) \tag{2.17}$$

During the calculations, a series of important integrals appear with the complex $z$ or real $|\,z\,|$ variables [10]:

($I_{BG}$) The *fundamental Barut-Girardello integral*:

$$\left(I_{BG}\right)\quad\int_0^{R\le\infty}d(|\,z\,|^2)(|\,z\,|^2)^n\,G_{p,q+1}^{q+1,0}\left(|\,z\,|^2\left|\begin{array}{ll}/\ ; & \{a_i-1\}_1^p\\ 0\,,\ \{b_j-1\}_1^q; & /\end{array}\right.\right)=\frac{\prod\limits_{j=1}^{q}\Gamma(b_j)}{\prod\limits_{i=1}^{p}\Gamma(a_i)}\rho_{p,q}(b/a\,|\,n) \tag{2.18}$$

Using the DOOT rules [10], we can write the BG-CSs as

$$| z >_{BG} = \frac{1}{\sqrt{{}_pF_q\left(\{a_i\}_1^p \; ; \; \{b_j\}_1^q ; | z |^2\right)}} \, {}_pF_q\left(\{a_i\}_1^p \; ; \; \{b_j\}_1^q ; z \hat{E}_+\right) | 0 > \tag{2.19}$$

($II_{BG}$) The *BG- integral in complex space*

$$\left(II_{BG}\right) \quad \int \frac{d^2 z}{\pi} G_{p,q+1}^{q+1,0}\left( | z |^2 \middle| \begin{matrix} / \; ; & \{a_i - 1\}_1^p \\ 0, \; \{b_j - 1\}_1^q \; ; & / \end{matrix} \right) \# \, {}_pF_q\left(\{a_i\}_1^p \; ; \; \{b_j\}_1^q ; z \hat{E}_+\right) \times$$

$$\times \, {}_pF_q\left(\{a_i\}_1^p \; ; \; \{b_j\}_1^q \; ; \; z^* \, \hat{E}_-\right) \# = \frac{\prod_{j=1}^{q} \Gamma\left(b_j\right)}{\prod_{i=1}^{p} \Gamma\left(a_i\right)} \; \# \, {}_pF_q\left(\{a_i\}_1^p \; ; \; \{b_j\}_1^q \; ; \; \hat{E}_+ \hat{E}_-\right) \# \tag{2.20}$$

($III_{BG}$) The *BG- angular integral*

$$\left(III_{BG}\right) \quad \int_0^{2\pi} \frac{d\varphi}{2\pi} \# \, {}_pF_q\left(\{a_i\}_1^p ; \{b_j\}_1^q ; z \hat{E}_+\right) {}_pF_q\left(\{a_i\}_1^p ; \{b_j\}_1^q ; z^* \, \hat{E}_-\right) \# =$$
$$= \# \, {}_{2p}F_{2q+1}\left(\{a_i\}_1^p , \{a_i\}_1^p ; 1, \; \{b_j\}_1^q , \{b_j\}_1^q ; \hat{E}_+ \hat{E}_- | z |^2\right) \# \tag{2.21}$$

($IV_{BG}$) The *BG- real integral*

$$\left(IV_{BG}\right) \quad \int_0^{\infty} d\left(| z |^2\right) G_{p,q+1}^{q+1,0}\left( | z |^2 \middle| \begin{matrix} / \; ; & \{a_i - 1\}_1^p \\ 0, \; \{b_j - 1\}_1^q \; ; & / \end{matrix} \right) \times$$

$$\times \, {}_{2p}F_{2q+1}\left(\{a_i\}_1^p , \{a_i\}_1^p ; 1, \; \{b_j\}_1^q , \{b_j\}_1^q ; \; \hat{E}_+ \hat{E}_- | z |^2\right) = \frac{\prod_{j=1}^{q} \Gamma\left(b_j\right)}{\prod_{i=1}^{p} \Gamma\left(a_i\right)} \; {}_pF_q\left(\{a_i\}_1^p \; ; \; \{b_j\}_1^q ; \hat{E}_+ \hat{E}_-\right) \tag{2.22}$$

***b) The Klauder-Perelomov coherent states (KP-CSs)*** are the result of action of the generalized displacement unitary operator $\# \exp\left(z \hat{E}_+ - z^* \hat{E}_-\right) \#$ on the vacuum state $| 0 >$ [6]:

$$| z >_{KP} = \frac{1}{\sqrt{{}_qF_p\left(\{b_j\}_1^q ; \{a_i\}_1^p ; | z |^2\right)}} \exp\left(z \hat{E}_+\right) | 0 > =$$
$$= \frac{1}{\sqrt{{}_qF_p\left(\{b_j\}_1^q ; \{a_i\}_1^p ; | z |^2\right)}} \sum_{n=0}^{M} \frac{\sqrt{\rho_{p,q}(b/a| n)}}{n!} z^n \, | n > \tag{2.23}$$

Comparing with the normalization function of the BG-CSs, it is to observe the interchanging of the sets of parameters $\{a_i\}_1^p$ and $\{b_j\}_1^q$. This is a manifestation of the duality between BG-CSs and KP-CSs. As will be seen below, the duality manifests itself through the mutual interchange, in the related functions and integrals, of the indices $p$ and $q$, as well as the sets of numbers $\{a_i\}_1^p$ and $\{b_j\}_1^q$. Only the projector of the vacuum state remains unchanged [15].

The corresponding integration measure for KP-CSs is

$$d\mu_{KP}(z) = \frac{\prod_{j=1}^{q}\Gamma(b_j)}{\prod_{i=1}^{p}\Gamma(a_i)} \frac{d\varphi}{2\pi} d(|z|^2)\; {}_qF_p\left(\{b_j\}_1^q;\{a_i\}_1^p;|z|^2\right) G_{q,p+1}^{p+1,0}\left( |z|^2 \,\middle|\, \begin{matrix} / \,; & \{b_j-1\}_1^q \\ 0,\{a_i-1\}_1^p\,; & / \end{matrix}\right) \quad (2.24)$$

($I_{BG}$) The *fundamental Klauder-Perelomov integral*:

$$(I_{KP})\quad \int_0^{R\le\infty} d(|z|^2)(|z|^2)^n G_{q,p+1}^{p+1,0}\left( |z|^2 \,\middle|\, \begin{matrix} / \,; & \{b_j-1\}_1^q \\ 0,\{a_i-1\}_1^p\,; & / \end{matrix}\right) = \frac{\prod_{i=1}^{p}\Gamma(a_i)}{\prod_{j=1}^{q}\Gamma(b_j)} \rho_{q,p}(a/b\,|\,n) \quad (2.25)$$

($II_{KP}$) The *KP- integral in complex space*

$$(II_{KP})\quad \int \frac{d^2z}{\pi} G_{q,p+1}^{p+1,0}\left( |z|^2 \,\middle|\, \begin{matrix} / \,; & \{b_j-1\}_1^q \\ 0,\; \{a_i-1\}_1^p\;; & / \end{matrix}\right) \#\exp\left(z\hat{E}_+\right)\exp\left(z^*\hat{E}_-\right)\# =$$

$$= \frac{\prod_{i=1}^{p}\Gamma(a_i)}{\prod_{j=1}^{q}\Gamma(b_j)} \#\, {}_pF_q\left(\{a_i\}_1^p;\{b_j\}_1^q;\hat{E}_+\hat{E}_-\right)\# \quad (2.26)$$

($III_{KP}$) The *KP- angular integral*

$$(III_{KP})\quad \int_0^{2\pi}\frac{d\varphi}{2\pi}\#\exp\left(z\hat{E}_+\right)\exp\left(z^*\hat{E}_-\right)\# = \#\,{}_0F_1\left(;1;\hat{E}_+\hat{E}_-|z|^2\right)\# = \# I_0\left(2|z|\sqrt{\hat{E}_+\hat{E}_-}\right)\# \quad (2.27)$$

($IV_{BG}$) The *KP- real integral*

$$(IV_{KP})\quad \int_0^\infty d\left(|z|^2\right) G_{q,p+1}^{p+1,0}\left( |z|^2 \,\middle|\, \begin{matrix} / \,; & \{b_j-1\}_1^q \\ 0,\; \{a_i-1\}_1^p\;; & / \end{matrix}\right) \# I_0\left(2|z|\sqrt{\hat{E}_+\hat{E}_-}\right)\# =$$

$$= \frac{\prod_{i=1}^{p}\Gamma(a_i)}{\prod_{j=1}^{q}\Gamma(b_j)} \#\,{}_pF_q\left(\{a_i\}_1^p;\{b_j\}_1^q;\hat{E}_+\hat{E}_-\right)\# \quad (2.28)$$

***c) The Gazeau-Klauder coherent states (GK-CSs)*** [4] have the following expansion:

$$| J,\gamma > = \frac{1}{\sqrt{{}_pF_q\left(\{a_i\}_1^p ; \{b_j\}_1^q ; J\right)}} \sum_{n=0}^{\infty} \frac{\left(\sqrt{J}\right)^n}{\sqrt{\rho_{p,q}(b/a\,|\,n)}} e^{-\mathbf{i}\gamma e(n)} \,| n > \tag{2.29}$$

where $0 \le J \le \infty$ is a real number labeling the GK-CSs and $-\infty \le \gamma \le +\infty$ is a real characteristic parameter.

The GK-CSs can be obtained if three steps are taken [13]:

1. We define BG-CSs, but for the *real* variable $J$, denoted by $| J >$:

$$\hat{E}_- \,| J > = J \,| J > \tag{2.30}$$

2. We develop $| J >$ into the Fock vectors base:

$$| J > = \frac{1}{\sqrt{{}_pF_q\left(\{a_i\}_1^p ; \{b_j\}_1^q ; J\right)}} \sum_{n=0}^{\infty} \frac{\left(\sqrt{J}\right)^n}{\sqrt{\rho_{p,q}(b/a\,|\,n)}} | n > \tag{2.31}$$

3. We act with the exponential operator $\exp\left(-\mathrm{i}\,\gamma\,\hat{H}\right)$ on the state $| J >$ (the parameter $\gamma$, as well as the Hamiltonian $\hat{H}$ are considered less dimensional):

$$| J,\gamma > = \exp\left(-\mathbf{i}\,\gamma\,\hat{H}\right) | J > = \frac{1}{\sqrt{{}_pF_q\left(\{a_i\}_1^p ; \{b_j\}_1^q ; J\right)}} \sum_{n=0}^{M} \frac{\left(\sqrt{J}\right)^n}{\sqrt{\rho_{p,q}(b/a\,|\,n)}} e^{-\mathbf{i}\gamma e(n)} \,| n > \tag{2.32}$$

Consequently, the integration measure for GK-CSs is

$$d\mu_{KP}(J,\gamma) = \frac{\prod_{i=1}^{p}\Gamma(a_i)}{\prod_{j=1}^{q}\Gamma(b_j)} \frac{d\gamma}{2R} dJ \;{}_pF_q\left(\{a_i\}_1^p ; \{b_j\}_1^q ; J\right) G_{p,q+1}^{q+1,0}\left( J \,\middle|\, \begin{matrix} / \,; & \{a_i - 1\}_1^p \\ 0,\ \{b_j - 1\}_1^m \,; & / \end{matrix} \right) \tag{2.33}$$

which is identical, from the mathematical point of view, with Eq. (2.16).

Consequently, the main formulas for GK-CSs will be identical, from a mathematical point of view, to the corresponding ones for BG-CSs and therefore we will not reproduce them.

## 3. New generalized integrals and integral representations

Comparing the integrals annotated with Latin letters (I-IV) for the two types of CSs, BG- and KP-, we will find that they are identical from a mathematical point of view. The difference consists only in the inversion of the indices $p$ and $q$, respectively of the sets of numbers $\{a_i\}_1^p$ and $\{b_j\}_1^q$. This is a consequence of the duality between the two types of CSs [15].

On the other hand, the formalism of the DOOT technique allows that the operators normally ordered under the DOOT sign # # can be treated as simple c-numbers (and, if necessary, they can be removed from the integration sign). This means that they can be replaced by simple non operatorial quantities. We will proceed as follows with the substitutions: $\hat{E}_+ \to A$ and $\hat{E}_- \to B$. Obviously, then the normal ordering DOOT sign $\# \ldots \#$ will also disappear.

In this sense, the main integrals resulting from the CSs formalism will be transcribed as follows:

(*I*) The *fundamental integrals*

$$(I_a) \quad \int_0^{R\le\infty} d(|z|^2)(|z|^2)^n G_{p,q+1}^{q+1,0}\left( |z|^2 \middle| \begin{matrix} / \,; & \{a_i-1\}_1^p \\ 0,\ \{b_j-1\}_1^q \,; & / \end{matrix} \right) = \frac{\prod_{j=1}^q \Gamma(b_j)}{\prod_{i=1}^p \Gamma(a_i)} \rho_{p,q}(b/a\,|\,n) \quad (3.1)$$

$$(I_b) \quad \int_0^{R\le\infty} d(|z|^2)(|z|^2)^n G_{q,p+1}^{p+1,0}\left( |z|^2 \middle| \begin{matrix} / \,; & \{b_j-1\}_1^q \\ 0,\ \{a_i-1\}_1^p \,; & / \end{matrix} \right) = \frac{\prod_{i=1}^p \Gamma(a_i)}{\prod_{j=1}^q \Gamma(b_j)} \frac{(n!)^2}{\rho_{p,q}(b/a\,|\,n)} \quad (3.2)$$

(*II*) The *integrals in complex space*

$$(II_a) \quad \int \frac{d^2z}{\pi} G_{p,q+1}^{q+1,0}\left( |z|^2 \middle| \begin{matrix} / \,; & \{a_i-1\}_1^p \\ 0,\ \{b_j-1\}_1^q \,; & / \end{matrix} \right) {}_pF_q\left(\{a_i\}_1^p;\{b_j\}_1^q;Az\right)\times$$

$$\times\ {}_pF_q\left(\{a_i\}_1^p;\{b_j\}_1^q;Bz^*\right) = \frac{\prod_{j=1}^q \Gamma(b_j)}{\prod_{i=1}^p \Gamma(a_i)}\, {}_pF_q\left(\{a_i\}_1^p;\{b_j\}_1^q;AB\right) \quad (3.3)$$

$$(II_b) \quad \int \frac{d^2z}{\pi} G_{q,p+1}^{p+1,0}\left( |z|^2 \middle| \begin{matrix} / \,; & \{b_j-1\}_1^q \\ 0,\ \{a_i-1\}_1^p \,; & / \end{matrix} \right) \exp(Az)\exp(Bz^*) =$$

$$= \frac{\prod_{i=1}^p \Gamma(a_i)}{\prod_{j=1}^q \Gamma(b_j)}\, {}_pF_q\left(\{a_i\}_1^p;\{b_j\}_1^q;AB\right) \quad (3.4)$$

(*III*) The *angular integrals*

$$(III_a) \quad \int_0^{2\pi} \frac{d\varphi}{2\pi} \, {}_pF_q\left(\{a_i\}_1^p ; \{b_j\}_1^q ; Az\right) {}_pF_q\left(\{a_i\}_1^p ; \{b_j\}_1^q ; Bz^*\right) = \\ = {}_{2p}F_{2q+1}\left(\{a_i\}_1^p , \{a_i\}_1^p ; 1, \{b_j\}_1^q , \{b_j\}_1^q ; AB|z|^2\right) \tag{3.5}$$

$$(III_b) \quad \int_0^{2\pi} \frac{d\varphi}{2\pi} \exp(Az)\exp(Bz^*) = {}_0F_1\left( ; 1 ; AB|z|^2\right) = I_0\left(2|z|\sqrt{AB}\right) \tag{3.6}$$

($IV$) The *integrals in real space*

$$(IV_a) \quad \int_0^{\infty} d\left(|z|^2\right) G_{p,q+1}^{q+1,0}\left( |z|^2 \middle| \begin{matrix} / \, ; & \{a_i - 1\}_1^p \\ 0, \; \{b_j - 1\}_1^q \; ; & / \end{matrix} \right) \times \\ \times {}_{2p}F_{2q+1}\left(\{a_i\}_1^p , \{a_i\}_1^p ; 1, \{b_j\}_1^q , \{b_j\}_1^q ; AB|z|^2\right) = \frac{\prod_{j=1}^{q} \Gamma(b_j)}{\prod_{i=1}^{p} \Gamma(a_i)} \, {}_pF_q\left(\{a_i\}_1^p ; \{b_j\}_1^q ; AB\right) \tag{3.7}$$

$$(IV_b) \quad \int_0^{\infty} d\left(|z|^2\right) G_{q,p+1}^{p+1,0}\left( |z|^2 \middle| \begin{matrix} / \, ; & \{b_j - 1\}_1^q \\ 0, \; \{a_i - 1\}_1^p \; ; & / \end{matrix} \right) I_0\left(2|z|\sqrt{AB}\right) = \\ = \frac{\prod_{i=1}^{p} \Gamma(a_i)}{\prod_{j=1}^{q} \Gamma(b_j)} \, {}_pF_q\left(\{a_i\}_1^p ; \{b_j\}_1^q ; AB\right) \tag{3.8}$$

If we consider another hypergeometric function ${}_rF_s\left(\{c_i\}_1^r ; \{d_j\}_1^s ; A|z|^2\right)$ and develop it in series according to the powers of $|z|^2$, we will obtain the following integral:

$$(I_{G\times F})_a \quad \int_0^{\infty} d\left(|z|^2\right) G_{p,q+1}^{q+1,0}\left( |z|^2 \middle| \begin{matrix} / \, ; & \{a_i - 1\}_1^p \\ 0, \; \{b_j - 1\}_1^q \; ; & / \end{matrix} \right) {}_rF_s\left(\{c_i\}_1^r ; \{d_j\}_1^s ; A|z|^2\right) = \\ = \frac{\prod_{j=1}^{q} \Gamma(b_j)}{\prod_{i=1}^{p} \Gamma(a_i)} \, {}_{q+r+1}F_{p+s}\left(\{b_j\}_1^q , \{c_i\}_1^r , 1 ; \{a_i\}_1^p , \{d_j\}_1^s ; A\right) \tag{3.9}$$

The corresponding integral that comes from KP-CSs is

$$\left(I_{G\times F}\right)_b \qquad \int_0^\infty d\left(|z|^2\right) G_{q,p+1}^{p+1,0}\left( |z|^2 \middle| \begin{matrix} / \,; & \{b_j-1\}_1^q \\ 0, \quad \{a_i-1\}_1^p \;; & / \end{matrix} \right) {}_rF_s\left(\{c_i\}_1^r ; \{d_j\}_1^s ; A|z|^2\right) =$$

$$= \frac{\prod_{i=1}^{p}\Gamma(a_i)}{\prod_{j=1}^{q}\Gamma(b_j)} {}_{p+r+1}F_{q+s}\left(\{a_i\}_1^p , \{c_i\}_1^r , 1 ; \{b_j\}_1^q , \{d_j\}_1^s ; A\right) \qquad (3.10)$$

To these new kind of integrals we can add also the integrals previously obtained, which can be considered as the particular cases of the previous:

$$\left(I_{G\times F}\right)_c \qquad \int_0^\infty d\left(|z|^2\right) G_{p,q+1}^{q+1,0}\left( \frac{1}{C}|z|^2 \middle| \begin{matrix} / \,; & \{a_i-1\}_1^p \\ 0, \quad \{b_j-1\}_1^q \;; & / \end{matrix} \right) \times$$

$$\times {}_{2p}F_{2q+1}\left(\{a_i\}_1^p , \{a_i\}_1^p ; 1, \{b_j\}_1^q , \{b_j\}_1^q ; A|z|^2\right) = \frac{\prod_{j=1}^{q}\Gamma(b_j)}{\prod_{i=1}^{p}\Gamma(a_i)} {}_pF_q\left(\{a_i\}_1^p ; \{b_j\}_1^q ; A\right) \qquad (3.11)$$

Checking its correctness can be done using the classical integrals from one or two Meijer's $G$ functions, as well as some of their properties [15]:

$$\int_0^\infty d\left(|z|^2\right)\left(|z|^2\right)^{s-1} G_{p,q}^{m,n}\left( C|z|^2 \middle| \begin{matrix} \{a_i\}_1^n \,; & \{a_i\}_{n+1}^p \\ \{b_j\}_1^m ; & \{b_j\}_{m+1}^q \end{matrix} \right) = \frac{1}{C^s} \frac{\prod_{j=1}^{m}\Gamma(b_j+s)\prod_{i=1}^{n}\Gamma(1-a_i-s)}{\prod_{i=n+1}^{p}\Gamma(a_i+s)\prod_{j=m+1}^{q}\Gamma(1-b_j-s)} \qquad (3.10)$$

$$\int_0^\infty d\left(|z|^2\right)\left(|z|^2\right)^{\alpha-1} G_{p,q}^{m,n}\left( |z|^2 \middle| \begin{matrix} \{a_i\}_1^n \,; & \{a_i\}_{n+1}^p \\ \{b_j\}_1^m ; & \{b_j\}_{m+1}^q \end{matrix} \right) G_{u,v}^{s,t}\left( -A|z|^2 \middle| \begin{matrix} \{c_i\}_1^t \,; & \{c_i\}_{t+1}^u \\ \{d_j\}_1^s ; & \{d_j\}_{s+1}^v \end{matrix} \right) =$$

$$= \frac{1}{(-A)^\alpha} G_{p+v,q+u}^{m+t,n+s}\left( -\frac{1}{A} \middle| \begin{matrix} \{a_i\}_1^n , \{1-\alpha-d_j\}_1^s ; & \{1-\alpha-d_j\}_{s+1}^v , & \{a_i\}_{n+1}^p \\ \{b_j\}_1^m , \{1-\alpha-c_i\}_1^t ; & \{1-\alpha-c_i\}_{t+1}^u , & \{b_j\}_{m+1}^q \end{matrix} \right) \qquad (3.11)$$

By particularizing the integer indexes $p$ and $q$, respectively the sets of numbers $\{a_i\}_1^p$ and $\{b_j\}_1^q$ we will obtain some new integrals in which Meijer's G and hypergeometric functions involve. As an immediate consequence, new integral representations of some polynomials and special functions will result.

## 4. Some examples of new integrals and integral representations

Before to particularize the integrals I-IV above, let's recall some particular expressions for hypergeometric functions, Meijer's G functions, as well as some known integrals, which will be useful in verifying the results obtained through the CS formalism and using the DOOT technique. These can be found on the Wolfram Functions website [16], [17].

$$ {}_0F_0(\ ;\ ;\ x)=e^{x} \quad , \qquad {}_1F_0(a;\ ;\ x)=(1-x)^{-a} \tag{4.1}$$

$$ {}_0F_1(\ ;b\ ;x)=\begin{cases}\Gamma(b)\left(\sqrt{-x}\right)^{1-b}J_{b-1}(-x)\\ \Gamma(b)\left(\sqrt{x}\right)^{1-b}I_{b-1}(x)\end{cases} \tag{4.2}$$

$$ {}_0F_1\left(\ ;\frac{3}{2};\begin{Bmatrix}-1\\+1\end{Bmatrix}\frac{x^2}{4}\right)=\begin{Bmatrix}\dfrac{\sin x}{x}\\ \dfrac{\sinh x}{x}\end{Bmatrix} \quad , \qquad {}_0F_1\left(\ ;\frac{1}{2};\begin{Bmatrix}-1\\+1\end{Bmatrix}\frac{x^2}{4}\right)=\begin{Bmatrix}\cos x\\ \cosh x\end{Bmatrix} \tag{4.3}$$

$$ {}_1F_1(-n\ ;1\ ;\ x)=L_n(x) \quad , \qquad {}_1F_1(-n\ ;\ \lambda+1\ ;\ x)=\frac{n!}{(\lambda+1)_n}L_n^{\lambda}(x) \tag{4.4}$$

$$ {}_2F_0(-n\ ,\ b\ ;\ ;\ x)=n!\,x^nL_n^{-b-n}\left(-\frac{1}{x}\right),\quad {}_2F_0(-n\ ,n+1;\ ;\ x)=\frac{1}{\sqrt{\pi}}\frac{1}{\sqrt{-x}}e^{-\frac{1}{2x}}K_{n+\frac{1}{2}}\left(-\frac{1}{2x}\right) \tag{4.5}$$

$$ {}_2F_0\left(-\frac{n}{2}\ ,\ \frac{1-n}{2}\ ;\ ;\ x\right)=\left(\frac{\sqrt{-x}}{2}\right)^{n}H_n\left(\frac{1}{\sqrt{-x}}\right),\ {}_2F_0\left(-\frac{n}{2}\ ,\frac{1-n}{2}\ ;\ ;\ -\frac{1}{x^2}\right)=(2x)^nH_n(x) \tag{4.6}$$

$$ G_{0,1}^{1,0}\left(x\,\middle|\,b\right)=e^{-x}x^{b} \quad , \quad G_{0,2}^{2,0}\left(x\,\middle|\,b_1\,,\,b_2\right)=2\left(\sqrt{x}\right)^{b_1+b_2}K_{b_1-b_2}\left(2\sqrt{x}\right) \tag{4.7}$$

***General case 1*** - $p=0\ ,\ q=0\ ;\ A=|A|>0\ ,\ B<0\ ,\ B=-|B|$, integral of kind $(II_b)$

$$ \int\frac{d^2z}{\pi}G_{0,1}^{1,0}\left(|z|^2\,|\,0\right)\exp\left(|A|z\right)\exp\left(-|B|z^*\right)={}_0F_0(\ ;\ ;-|AB|) \tag{4.8}$$

The proof is simple and we obtain a well-known relation:

$$ \int\frac{d^2z}{\pi}\exp\left(-|z|^2\right)\exp\left(|A|z\right)\exp\left(-|B|z^*\right)\exp\left(-|A||B|\right)=\int\frac{d^2z}{\pi}\exp\left[-\left(z-|B|\right)\left(z^*-|A|\right)\right]=1 \tag{4.9}$$

***Example 1.1*** - $p=1\ ,\ q=0\ ;\ a_1=a\ ,\ A,\ B-$arbitrarily, integral of kind $(II_a)$.

$$\int \frac{d^2 z}{\pi} G_{1,1}^{1,0}\left( |z|^2 \middle| \begin{matrix} / ; & a-1 \\ 0 ; & / \end{matrix} \right) {}_1F_0\left(a ; ; Az\right) {}_1F_0\left(a ; ; Bz^*\right) = \frac{1}{\Gamma(a)} {}_1F_0\left(a ; ; AB\right) = \frac{1}{\Gamma(a)} \frac{1}{(1-AB)^a} \tag{4.10}$$

***General case 2 -*** $p=1$ , $q=0$ ; $a_1 = a$ , $AB = x$, integral of kind $(IV_a)$ .

$$\int_0^\infty d\left(|z|^2\right) G_{1,1}^{1,0}\left( |z|^2 \middle| \begin{matrix} / ; & a-1 \\ 0 ; & / \end{matrix} \right) {}_2F_1\left(a, a ; 1; x|z|^2\right) = \Gamma(a)\, {}_1F_0\left(a ; ; x\right) = \frac{1}{\Gamma(a)} \frac{1}{(1-x)^a} \tag{4.11}$$

For the proof, first the hypergeometric function ${}_2F_1(...)$ in power series will be developed.

***General case 3 -*** $p=0$ , $q=1$ ; $b_1 = \frac{m}{2}$ , $AB = f(x)$, $m$ - natural number, integral of kind $(IV_a)$ .

$$\int_0^\infty d\left(|z|^2\right) G_{0,2}^{2,0}\left( |z|^2 \middle| 0, \ \frac{m}{2}-1 \right) {}_0F_3\left( \ ; 1, \ \frac{m}{2}, \frac{m}{2}; \ f(x)|z|^2 \right) = \Gamma\left(\frac{m}{2}\right) {}_0F_1\left( \ ; \frac{m}{2}; f(x) \right) \tag{4.12}$$

For proof, first develop ${}_0F_3(...)$ in power series and then use Eq. (4.7) and the following integral [18]:

$$\int_0^\infty dt\, t^{\alpha-1} K_\nu(t) = 2^{\alpha-2}\, \Gamma\left(\frac{\alpha-\nu}{2}\right) \Gamma\left(\frac{\alpha+\nu}{2}\right) \quad , \quad \mathrm{Re}\,\alpha > |\mathrm{Re}\,\nu| \tag{4.13}$$

From Eq. (4.12), we will obtain new integral representations for trigonometric and hyperbolic functions.

***Example 3.1 -*** $m=3$ , $\Gamma\left(\frac{3}{2}\right) = \frac{\sqrt{\pi}}{2}$ , $f(x) = \pm\frac{x^2}{4}$

$$\begin{Bmatrix} \dfrac{\sin x}{x} \\ \dfrac{\sinh x}{x} \end{Bmatrix} = \frac{2}{\sqrt{\pi}} \int_0^\infty d\left(|z|^2\right) G_{0,2}^{2,0}\left( |z|^2 \middle| 0, \ \frac{1}{2} \right) {}_0F_3\left( \ ; 1, \ \frac{3}{2}, \frac{3}{2}; \begin{Bmatrix} -1 \\ +1 \end{Bmatrix} \frac{x^2}{4} |z|^2 \right) \tag{4.14}$$

***Example 3.2 -*** $m=1$ , $\Gamma\left(\frac{1}{2}\right) = \sqrt{\pi}$ , $f(x) = \pm\frac{x^2}{4}$

$$\left.\begin{cases}\cos x\\ \\ \cosh x\end{cases}\right\}=\frac{1}{\sqrt{\pi}}\int_0^\infty d\left(|z|^2\right)G_{0,2}^{2,0}\left(|z|^2\middle|0,\quad -\frac{1}{2}\right){}_0F_3\left(\quad;1,\ \frac{1}{2},\frac{1}{2};\begin{Bmatrix}-1\\+1\end{Bmatrix}\frac{x^2}{4}|z|^2\right)\tag{4.14}$$

***Example 3.3 -*** $p=0$ , $q=1$ ; $b_1=\dfrac{m}{2}$ , $f(x)=\begin{Bmatrix}\mathrm{i}\\1\end{Bmatrix}\dfrac{x^2}{4}$, of kind (3.8) $(IV_b)$.

$$\int_0^\infty d\left(|z|^2\right)G_{1,1}^{1,0}\left(|z|^2\middle|\begin{matrix}/\ ; & \frac{m}{2}-1_1\\ 0\ ; & /\end{matrix}\right)I_0\left(2|z|\sqrt{\begin{Bmatrix}\mathrm{i}\\1\end{Bmatrix}\frac{x^2}{4}}\right)=\frac{1}{\Gamma(b_1)}{}_0F_1\left(;\frac{m}{2};\begin{Bmatrix}\mathrm{i}\\1\end{Bmatrix}\frac{x^2}{4}\right)\tag{4.15}$$

It is well known that Laguerre's generalized polynomials are related to certain hypergeometric functions:

$$L_n^\lambda(x)=\begin{cases}\dfrac{(\lambda+1)_n}{n\,!}e^{-x}\ {}_1F_1(\lambda+1+n\,;\lambda+1;-x)\\ \dfrac{(\lambda+1)_n}{n\,!}\ {}_1F_1(-n\,;\lambda+1;\,x)\\ \dfrac{1}{n\,!}\dfrac{1}{x^n}\ {}_2F_0\left(-n\,,-\lambda+n\,;-\dfrac{1}{x}\right)\end{cases}\tag{4.16}$$

Let's derive some new representations of Laguerre's generalized polynomials.

***General case 4 -*** $p=1$ , $q=1$, $a_1=a$ , $b_1=b$ , $AB=f(x)$

$$\int_0^\infty d\left(|z|^2\right)G_{1,2}^{2,0}\left(|z|^2\middle|\begin{matrix}/\ ; & b\\ 0,\,a-1; & /\end{matrix}\right)I_0\left(2|z|\sqrt{f(x)}\right)=\frac{\Gamma(a)}{\Gamma(b)}{}_1F_1(a\,;b\,;\,f(x))\tag{4.17}$$

***Example 4.1 -*** $p=1$ , $q=1$, $a_1=-n$ , $b_1=\lambda+1$ , $AB=x$

$$\begin{aligned}&\int_0^\infty d\left(|z|^2\right)G_{1,2}^{2,0}\left(|z|^2\middle|\begin{matrix}/\ ; & \lambda\\ 0,\,-n-1\,; & /\end{matrix}\right)I_0\left(2|z|\sqrt{x}\right)=\\ &=\frac{\Gamma(-n)}{\Gamma(\lambda+1)}{}_1F_1(-n;\,\lambda+1;x)=\frac{\Gamma(-n)\qquad n!}{\Gamma(\lambda+1)(\lambda+1)_n}L_n^\lambda(x)\end{aligned}\tag{4.18}$$

from which a new representation for Laquerre's generalized polynomials results:

$$L_n^\lambda(x)=\frac{1}{\Gamma(-n)}\frac{\Gamma(\lambda+1+n)}{n\,!}\int_0^\infty d\left(|z|^2\right)G_{1,2}^{2,0}\left(|z|^2\middle|\begin{matrix}/\ ; & \lambda\\ 0,\ \ -n-1\ ; & /\end{matrix}\right)I_0\left(2|z|\sqrt{x}\right)\tag{4.19}$$

As a particular case, if $\lambda=0$, we get

$$L_n(x)=\frac{1}{\Gamma(-n)}\int_0^\infty d\left(|z|^2\right)G_{0,1}^{1,0}\left(|z|^2\,|-n-1\right)I_0\left(2|z|\sqrt{x}\right) \tag{4.20}$$

In order to obtain the Laguerre polynomials with negative argument we examine the following case:

***Example 4.2 -*** $p=1$ , $q=1$, $a_1=a$ , $b_1=a-n$ , $AB=x$

$$\begin{aligned}&\int_0^\infty d\left(|z|^2\right)G_{1,2}^{2,0}\left(|z|^2\left|\begin{matrix}/\,; & a-n-1\\ 0,\ a-1\,; & /\end{matrix}\right.\right)I_0\left(2|z|\sqrt{x}\right)=\\ &=\frac{\Gamma(a)}{\Gamma(a-n)}\,{}_1F_1(a;\,a-n;x)=\frac{\Gamma(a)\quad(-1)^n n!}{\Gamma(a-n)(1-a)_n}e^x\,L_n^{a-n-1}(-x)\end{aligned} \tag{4.21}$$

where we used the following equation involving the Tricomi confluent hypergeometric function $U(a,b,t)$ [18] :

$$\int_0^\infty dt\,e^{-t}\,t^{\alpha-1}U(a,b,t)=\frac{\Gamma(1-b+\alpha)\Gamma(\alpha)}{\Gamma(a-b+\alpha+1)}\ ,\quad \max\{\mathrm{Re}(b)-1\}<\mathrm{Re}(\alpha) \tag{4.22}$$

as wel as the relation involving the Pochhammer symbol $(1-a)_n=(-1)^n\,\dfrac{\Gamma(a)}{\Gamma(a-n)}$ .

$$L_n^{a-n-1}(-x)=\frac{1}{n\,!}\,e^{-x}\int_0^\infty d\left(|z|^2\right)G_{1,2}^{2,0}\left(|z|^2\left|\begin{matrix}/\,; & a-n-1\\ 0,\ a-1\,; & /\end{matrix}\right.\right)I_0\left(2|z|\sqrt{x}\right) \tag{4.23}$$

Particularly, for $a=n+1$, we obtain another integral representation for Laguerre polynomial with negative argument.

$$L_n(-x)=\frac{1}{n\,!}\,e^{-x}\int_0^\infty d\left(|z|^2\right)G_{0,1}^{1,0}\left(|z|^2\,|\ n\ \right)I_0\left(2|z|\sqrt{x}\right)={}_1F_1(n+1;1;x) \tag{4.24}$$

Another integral representation of the generalized Laguerre polynomials results from the following case:

***Example 4.3 -*** $p=2$ , $q=0$, $a_1=-n$ , $a_2=a$, $AB=-\dfrac{1}{x}$ $(IV_b)$

$$\begin{aligned}&\int_0^\infty d\left(|z|^2\right)G_{0,3}^{3,0}\left(|z|^2\,|\,0\ ,\ -n-1\ ,\ a-1\right)I_0\left(2|z|\sqrt{-\frac{1}{x}}\right)=\\ &=\Gamma(-n)\Gamma(a)\ {}_2F_0\left(-n;\,a;-\frac{1}{x}\right)=\Gamma(-n)\Gamma(a)\,n!\left(-\frac{1}{x}\right)^n L_n^{-a-n}(x)\end{aligned} \tag{4.25}$$

$$L_n^{-a-n}(x)=(-1)^n \frac{1}{\Gamma(-n)\Gamma(a)} \frac{x^n}{n!} \int_0^\infty d(|z|^2)\, G_{0,3}^{3,0}(|z|^2 \,|\, 0\,,\ -n-1\,,\ a-1) I_0\left(2|z|\sqrt{-\frac{1}{x}}\right) \qquad (4.26)$$

For the proofs, wherever Meijer's G functions $G_{0,3}^{3,0}(|z|^2|...)$, or $G_{1,3}^{3,0}(|z|^2|...)$ appears, the Bessel function will be developed in series and the generalized integral for Meijer's G functions (3.10) will be used.

Particularly, for $a_2 = a = -n$, we obtain

$$L_n(x)=(-1)^n \frac{1}{[\Gamma(-n)]^2} \frac{x^n}{n!} \int_0^\infty d(|z|^2)\, G_{0,3}^{3,0}(|z|^2 \,|\, 0\,,\ -n-1\,,\ -n-1) I_0\left(2|z|\sqrt{-\frac{1}{x}}\right) \qquad (4.27)$$

For Hermite polynomials

$$H_n(x)=(2x)^n\ {}_2F_0\left(-\frac{n}{2};\frac{1-n}{2};-\frac{1}{x^2}\right) \qquad (4.28)$$

we have:

***General case 5 -*** $p=2\ ,\ q=0,\quad AB=f(x)\quad (IV_b)$

$$\int_0^\infty d(|z|^2)\, G_{0,3}^{3,0}(|z|^2 \,|\, 0\,,\ a_1-1\,,\ a_2-1) I_0\left(2|z|\sqrt{f(x)}\right)=\Gamma(a_1)\Gamma(a_2)\ {}_2F_0(a_1;a_2;f(x)) \qquad (4.29)$$

***Example 5.1 -*** $p=2\ ,\ q=0,\quad a_1=-\frac{n}{2}\ ,\ a_2=\frac{1-n}{2},\ f(x)=-\frac{1}{x^2}\quad (IV_b)$

$$\begin{aligned}&\int_0^\infty d(|z|^2)\, G_{0,3}^{3,0}\left(|z|^2 \middle|\, 0\,,\ -\frac{n}{2}-1\,,\ \frac{1-n}{2}-1\right) I_0\left(2|z|\sqrt{-\frac{1}{x^2}}\right)=\\ &=\Gamma\left(-\frac{n}{2}\right)\Gamma\left(\frac{1-n}{2}\right)\ {}_2F_0\left(-\frac{n}{2};\frac{1-n}{2};-\frac{1}{x^2}\right)=\Gamma\left(-\frac{n}{2}\right)\Gamma\left(\frac{1-n}{2}\right)\left(\frac{1}{2x}\right)^n H_n(x)\end{aligned} \qquad (4.30)$$

$$H_n(x)=\frac{(2x)^n}{\Gamma\left(-\frac{n}{2}\right)\Gamma\left(\frac{1-n}{2}\right)} \int_0^\infty d(|z|^2)\, G_{0,3}^{3,0}\left(|z|^2 \middle|\, 0\,,\ -\frac{n}{2}-1\,,\ \frac{1-n}{2}-1\right) I_0\left(2|z|\sqrt{-\frac{1}{x^2}}\right) \qquad (4.31)$$

Analogously, for the Bessel functions of the second kind

***Example 5.2 -*** $p=2\ ,\ q=0,\quad a_1=-n\ ,\ a_2=n+1,\ f(x)=-\frac{1}{2x}\quad (IV_b)$

$$\int_0^\infty d\left(|z|^2\right) G_{0,3}^{3,0}\left(|z|^2 \,|\, 0\,,\ -n-1\,,\ n\right) I_0\left(2|z|\sqrt{-\frac{1}{2x}}\right)=$$
$$=\Gamma(-n)\Gamma(n+1)\ {}_2F_0\left(-n;\ n+1;-\frac{1}{2x}\right)=\Gamma(-n)\Gamma(n+1)\frac{1}{\sqrt{\pi}}\sqrt{2x}\,e^{-x}K_{n+\frac{1}{2}}(x) \tag{4.32}$$

$$K_{n+\frac{1}{2}}(x)=\sqrt{\pi}\,\frac{1}{\Gamma(-n)n\,!}\frac{e^x}{\sqrt{2x}}\int_0^\infty d\left(|z|^2\right) G_{0,3}^{3,0}\left(|z|^2 \,|\, 0\,,\ -n-1\,,\ n\right) I_0\left(2|z|\sqrt{-\frac{1}{2x}}\right) \tag{4.33}$$

For the case $p=2$ , $q=1$ let's present the corresponding integral where $AB=f(x)$, is a certain function. For the different values of the parameters $a_1$, $a_2$ and $b_1$ and of this function, different integral representations are obtained for the orthogonal polynomials.

***General case 6 -*** $p=1$ , $q=1$, $a_1=a$ , $b_1=2a$, $AB=x$

$$\int_0^\infty d\left(|z|^2\right) G_{1,3}^{3,0}\left(|z|^2 \left|\begin{matrix} / \ ; & b_1-1 \\ 0,\ a_1-1\,,\ a_2-1\ ; & / \end{matrix}\right.\right) I_0\left(2|z|\sqrt{f(x)}\right)=$$
$$=\frac{\Gamma(a_1)\Gamma(a_2)}{\Gamma(b_1)}\ {}_2F_1\left(a_1\ ,\ a_2\ ;b_1\ ;f(x)\right) \tag{4.34}$$

By particularizing the parameters and the function $f(x)$ above, we will obtain 3 cases of integral representations for Legendre's polynomials. This is based on the following representations of Legendre's polynomials:

$$P_n(x)=\begin{cases} {}_2F_1\left(-n\ ,\ n+1\,;1\,;\dfrac{1-x}{2}\right) \\ \dfrac{(1+x)^n}{2^n}\ {}_2F_1\left(-n\ ,\ -n\,;1\ ;-\dfrac{1-x}{1+x}\right) \\ \dfrac{1}{\sqrt{\pi}}\dfrac{2^n\Gamma\left(n+\frac{1}{2}\right)}{n\,!}(x-1)^n\ {}_2F_1\left(-n\ ,\ -n\,;-2n\ ;\dfrac{2}{1-x}\right) \end{cases} \tag{4.35}$$

***Example 6.1 -*** $p=2$ , $q=1$ , $a_1=-n$ , $a_2=-n$ , $b_1=-2n$, $f(x)=\dfrac{2}{1-x}$

$$(IV_b)\quad \int_0^\infty d\left(|z|^2\right) G_{1,3}^{3,0}\left(|z|^2 \left|\begin{matrix} / \ ; & -2n-1 \\ 0,-n-1\,,-n-1\ ; & / \end{matrix}\right.\right) I_0\left(2|z|\sqrt{\frac{2}{1-x}}\right)=$$
$$=\frac{[\Gamma(-n)]^2}{\Gamma(-2n)}\ {}_2F_1\left(-n\ ,\ -n\,;-2n\ ;\frac{2}{1-x}\right)=\frac{[\Gamma(-n)]^2}{\Gamma(-2n)}\sqrt{\pi}\,2^{-n}\frac{n!}{\Gamma\left(n+\frac{1}{2}\right)}(x-1)^{-n}P_n(x) \tag{4.36}$$

$$P_n(x)=\frac{1}{\sqrt{\pi}}2^n\,\frac{\Gamma\left(n+\frac{1}{2}\right)}{n!}\,\frac{\Gamma(-2n)}{[\Gamma(-n)]^2}(x-1)^n\times$$
$$\times\int_0^\infty d\left(|z|^2\right)G_{1,3}^{3,0}\left(|z|^2\left|\begin{matrix}/\,; & -2n-1\\ 0,-n-1,-n-1\,; & /\end{matrix}\right.\right)I_0\left(2|z|\sqrt{\frac{2}{1-x}}\right)= \tag{4.37}$$

***Example 6.2 -*** $p=2\ ,\ \ q=1\ ,\ \ a_1=-n\ ,\ \ a_2=n+1\ \ ,\ b_1=1,\ \ f(x)=\dfrac{1-x}{2}$

$$(IV_b)\quad \int_0^\infty d\left(|z|^2\right)G_{1,3}^{3,0}\left(|z|^2\left|\begin{matrix}/\,; & 0\\ 0,-n-1,n\,; & /\end{matrix}\right.\right)I_0\left(2|z|\sqrt{\frac{1-x}{2}}\right)=$$
$$=\Gamma(-n)\,n\,!\;{}_2F_1\left(-n\ ,\ n\,;1\ ;\frac{1-x}{2}\right)=\Gamma(-n)\,n\,!\,P_n(x) \tag{4.38}$$

$$P_n(x)=\frac{1}{\Gamma(-n)\,n\,!}\int_0^\infty d\left(|z|^2\right)G_{1,3}^{3,0}\left(|z|^2\left|\begin{matrix}/\,; & 0\\ 0,-n-1,n\,; & /\end{matrix}\right.\right)I_0\left(2|z|\sqrt{\frac{1-x}{2}}\right) \tag{4.39}$$

During the deduction of this integral representation, the following representation also appears:

$$P_n(x)=\frac{2}{\Gamma(-n)\,n\,!}\int_0^\infty d\left(2|z|\right)K_{2n+1}(2|z|)I_0\left(2|z|\sqrt{\frac{1-x}{2}}\right) \tag{4.40}$$

which is in full agreement with 6.576.5, page 684 of the Gradshteyn - Ryzhik book [19].

***Example 6.3 -*** $p=2\ ,\ \ q=1\ ,\ \ a_1=-n\ ,\ \ a_2=-n\ \ ,\ b_1=1,\ \ f(x)=\dfrac{x-1}{x+1}$

$$(IV_b)\quad \int_0^\infty d\left(|z|^2\right)G_{1,3}^{3,0}\left(|z|^2\left|\begin{matrix}/\,; & 0\\ 0,-n-1,-n-1\,; & /\end{matrix}\right.\right)I_0\left(2|z|\sqrt{\frac{x-1}{x+1}}\right)=$$
$$=[\Gamma(-n)]^2\;{}_2F_1\left(-n\ ,\ -n\,;1\ ;\frac{x-1}{x+1}\right)=[\Gamma(-n)]^2\left(\frac{2}{x+1}\right)^n P_n(x) \tag{4.41}$$

$$P_n(x)=\frac{1}{[\Gamma(-n)]^2}\left(\frac{x+1}{2}\right)^n\int_0^\infty d\left(|z|^2\right)G_{1,3}^{3,0}\left(|z|^2\left|\begin{matrix}/\,; & 0\\ 0,-n-1,-n-1\,; & /\end{matrix}\right.\right)I_0\left(2|z|\sqrt{\frac{x-1}{x+1}}\right) \tag{4.42}$$

Analogously, for the Chebysev T polynomial of second kind, which is also expressible by hypergeometric functions

$$T_n(x)=\begin{cases} {}_2F_1\left(-n\,,\ n\,;\frac{1}{2}\ ;\frac{1-x}{2}\right) \\ (-1)^n\,{}_2F_1\left(-n\,,\ n\,;\frac{1}{2}\ ;\frac{1+x}{2}\right)\end{cases} \tag{4.43}$$

we will have the following new integral representations:

***General case 7 -*** $p=2\ ,\ \ q=1\ ,\ \ \ a_1=-n\ ,\ a_2=+n\ \ ,\ b_1=\frac{1}{2},\ \ \ f(x)=\frac{1\mp x}{2}$

$$\begin{aligned}(IV_b)\ \ &\int_0^\infty d\left(|z|^2\right)G_{1,3}^{3,0}\left(|z|^2\left|\begin{matrix}/\ ; & -\frac{1}{2}\\ 0,-n-1\,,n-1\ ; & /\end{matrix}\right.\right)I_0\left(2|z|\sqrt{\frac{1\mp x}{2}}\right)=\\ &=\frac{\Gamma(-n)\Gamma(n)}{\Gamma\left(\frac{1}{2}\right)}\,{}_2F_1\left(-n\,,\ n\,;\frac{1}{2}\ ;\frac{1\mp x}{2}\right)=\frac{\Gamma(-n)\Gamma(n)}{\Gamma\left(\frac{1}{2}\right)}\begin{Bmatrix}1\\(-1)^n\end{Bmatrix}T_n(x)\end{aligned} \tag{4.44}$$

$$T_n(x)=\begin{Bmatrix}1\\(-1)^n\end{Bmatrix}\frac{\sqrt{\pi}}{\Gamma(-n)\Gamma(n)}\int_0^\infty d\left(|z|^2\right)G_{1,3}^{3,0}\left(|z|^2\left|\begin{matrix}/\ ; & -\frac{1}{2}\\ 0,-n-1\,,n-1\ ; & /\end{matrix}\right.\right)I_0\left(2|z|\sqrt{\frac{1\mp x}{2}}\right) \tag{4.45}$$

Same thing for Chebysev U polynomial of second kind

$$U_n(x)=\begin{cases}(n+1)\,{}_2F_1\left(-n\,,\ n+2\,;\frac{3}{2}\ ;\frac{1-x}{2}\right)\\ (-1)^n\,(n+1)\,{}_2F_1\left(-n\,,\ n+2\,;\frac{3}{2}\ ;\frac{1+x}{2}\right)\end{cases} \tag{4.46}$$

***General case 8 -*** $p=2\ ,\ \ q=1\ ,\ \ \ a_1=-n\ ,\ a_2=n+2\ \ ,\ b_1=\frac{3}{2},\ \ \ f(x)=\frac{1\mp x}{2}$

$$\begin{aligned}(IV_b)\ \ &\int_0^\infty d\left(|z|^2\right)G_{1,3}^{3,0}\left(|z|^2\left|\begin{matrix}/\ ; & \frac{1}{2}\\ 0,-n-1\,,n+1\ ; & /\end{matrix}\right.\right)I_0\left(2|z|\sqrt{\frac{1\mp x}{2}}\right)=\\ &=\frac{\Gamma(-n)\Gamma(n+2)}{\Gamma\left(\frac{3}{2}\right)}\,{}_2F_1\left(-n\,,\ n+2\,;\frac{3}{2}\ ;\frac{1\mp x}{2}\right)=\frac{\Gamma(-n)\Gamma(n+2)}{\Gamma\left(\frac{3}{2}\right)}\frac{1}{n+1}\begin{Bmatrix}1\\(-1)^n\end{Bmatrix}U_n(x)\end{aligned} \tag{4.47}$$

$$U_n(x)=\begin{Bmatrix}1\\(-1)^n\end{Bmatrix}\frac{1}{2}\frac{\sqrt{\pi}}{\Gamma(-n)n!}\int_0^\infty d\left(|z|^2\right)G_{1,3}^{3,0}\left(|z|^2\left|\begin{matrix}/\ ; & \frac{1}{2}\\ 0,-n-1\,,n+1\ ; & /\end{matrix}\right.\right)I_0\left(2|z|\sqrt{\frac{1\mp x}{2}}\right) \tag{4.48}$$

For Bessel functions of the first kind the calculations are similar.

***General case 9 -*** $p=1$ , $q=1$, $a_1 = a$ , $b_1 = 2a$, $AB = f(x)$

$$\int_0^\infty d\left(|z|^2\right) G_{1,2}^{2,0}\left( |z|^2 \middle| \begin{array}{cc} / \,; & 2a-1 \\ 0, \; a-1 \,; & / \end{array} \right) I_0\left(2|z|\sqrt{f(x)}\right) = \frac{\Gamma(a)}{\Gamma(2a)} \, {}_1F_1\left(a; 2a; f(x)\right) \quad (4.49)$$

Let us we particularize the above integral for the cases of Bessel functions of the first kind $J_\nu(x)$ , and respectively the modified Bessel functions of the first kind $I_\nu(x)$.

If $a = \nu + \frac{1}{2}$ , and $AB = 2\left\{\begin{array}{c} \mathrm{i} \\ 1 \end{array}\right\} x$ we have

$$\left\{\begin{array}{c} J_\nu(x) \\ I_\nu(x) \end{array}\right\} = \frac{\exp\left(-\left\{\begin{array}{c} \mathrm{i} \\ 1 \end{array}\right\} x\right)}{\Gamma(\nu+1)} \left(\frac{x}{2}\right)^\nu \, {}_1F_1\left(\nu + \frac{1}{2}; 2\nu+1; 2\left\{\begin{array}{c} \mathrm{i} \\ 1 \end{array}\right\} x\right) \quad (4.50)$$

$$\int_0^\infty d\left(|z|^2\right) G_{1,2}^{2,0}\left( |z|^2 \middle| \begin{array}{cc} / \,; & 2\nu \\ 0, \; \nu - \frac{1}{2} \,; & / \end{array} \right) I_0\left(2|z|\sqrt{2\left\{\begin{array}{c} \mathrm{i} \\ 1 \end{array}\right\} x}\right) =$$

$$= \frac{\Gamma\left(\nu + \frac{1}{2}\right)}{\Gamma(2\nu+1)} \, {}_1F_1\left(\nu + \frac{1}{2}\,; 2\nu+1; 2\left\{\begin{array}{c} \mathrm{i} \\ 1 \end{array}\right\} x\right) = \frac{\Gamma\left(\nu + \frac{1}{2}\right)\Gamma(\nu+1)}{\Gamma(2\nu+1)} \exp\left(+\left\{\begin{array}{c} \mathrm{i} \\ 1 \end{array}\right\} x\right) \left(\frac{x}{2}\right)^{-\nu} \left\{\begin{array}{c} J_\nu(x) \\ I_\nu(x) \end{array}\right\} \quad (4.51)$$

$$\left\{\begin{array}{c} J_\nu(x) \\ I_\nu(x) \end{array}\right\} = \frac{1}{\sqrt{\pi}} \exp\left(-\left\{\begin{array}{c} \mathrm{i} \\ 1 \end{array}\right\} x\right) (2x)^\nu \int_0^\infty d\left(|z|^2\right) G_{1,2}^{2,0}\left( |z|^2 \middle| \begin{array}{cc} / \,; & 2\nu \\ 0, \; \nu - \frac{1}{2} \,; & / \end{array} \right) I_0\left(2|z|\sqrt{2\left\{\begin{array}{c} \mathrm{i} \\ 1 \end{array}\right\} x}\right) \quad (4.52)$$

in concordance with Eq. 8.335.1, page 896 of Gradshteyn and Ryshik's book [19]:

Let's turn our attention to the integrals which we are called the *foundamental CSs integrals.*

***General case 9 -*** $p=1$ , $q=1$, $a_1 = a$ , $b_1 = b$, $AB = f(x)$

***Example 9.1 -*** For $p=1$ and $q=1$, as well as $a_1 = 1$ , we have

$$G_{1,2}^{2,0}\left( |z|^2 \middle| \begin{array}{cc} / \,; & 0 \\ 0, \; b-1 \,; & / \end{array} \right) = G_{0,1}^{1,0}\left(|z|^2 \,\middle|\, b-1\right) = \left(|z|^2\right)^{b-1} e^{-|z|^2} \quad (4.52)$$

The integral (3.10) is then

$$\int_0^\infty d\left(|z|^2\right) G_{1,2}^{2,0}\left( |z|^2 \middle| \begin{matrix} / \,; & 0 \\ 0,\ \ b-1\,; & / \end{matrix} \right) {}_rF_s\left(\{c_i\}_1^r\,;\,\{d_j\}_1^s\,;\,f(x)|z|^2\right)= \\ =\Gamma(b)\,{}_{r+1}F_s\left(b,\{c_i\}_1^r\,;\,\{d_j\}_1^s\,;\,f(x)\right) \tag{4.53}$$

This result is in accordance with the integral 752.5, pp. 814 of from Gradshteyn and Ryshik's book [19]:

$$\int_0^\infty dx\, e^{-x} x^{s-1}\,{}_pF_q\left(a_1,a_2,...,a_p\,;\,b_1,b_2,...,b_q\,;A|z|^2\right)= \\ =\Gamma(s)\,{}_{p+1}F_q\left(s,a_1,a_2,...,a_p\,;\,b_1,b_2,...,b_q\,;\,A\right) \tag{4.54}$$

On the other hand this equation can be regarded as the Laplace transform of the generalized hypergeometric function [20].

***Example 9.2*** **-** Let's we choose $p=1$, $q=1$, $m=2$, $n=0$, and $a_1\ ,b_1\neq 0$. Then

$$G_{1,2}^{2,0}\left( |z|^2 \middle| \begin{matrix} / \,; & a_1-1 \\ 0,\ \ b_1-1\,; & / \end{matrix} \right) = e^{-|z|^2}U\left(a_1-b_1\,,2-b_1\,;|z|^2\right) \tag{4.55}$$

where $U\left(a_1-b_1\,,2-b_1\,;|z|^2\right)$ is the Tricomi confluent hypergeometric function. According to Eq. (3.10) we have

$$\int_0^\infty d\left(|z|^2\right) G_{1,2}^{2,0}\left( |z|^2 \middle| \begin{matrix} / \,; & a_1-1 \\ 0,\ \ b_1-1\,; & / \end{matrix} \right) {}_rF_s\left(\{c_i\}_1^r\,;\,\{d_j\}_1^s\,;\,f(x)|z|^2\right)= \\ =\Gamma(b_1)\,{}_{r+2}F_{s+1}\left(b_1\,,\{c_i\}_1^r\,,1;a_1\,,\{d_j\}_1^s\,;\,f(x)\right) \tag{4.56}$$

result which can be verified using the integral [20]:

$$\int_0^\infty d\left(|z|^2\right)\left(|z|^2\right)^{\alpha-1} e^{-|z|^2}U\left(a\,,b\,,|z|^2\right)=\frac{\Gamma(1-b+\alpha)\Gamma(\alpha)}{\Gamma(a-b+\alpha+1)}\,,\ \max\left(0,\mathrm{Re}(b)-1\right)<\mathrm{Re}(\alpha) \tag{4.57}$$

Generally, by particularizing all coefficients $p$, $q$, $r$ and $s$, as well as the set of parameters $\{a_i\}_1^p$, $\{b_j\}_1^q$, $\{c_i\}_1^r$ and $\{d_j\}_1^s$, it is possible to obtain a lot of new integrals (some of them unknown to yet).

Returning to the fundamental integral (2.18)

$$\left(I_{BG}\right)\int_0^{R\le\infty} d(|z|^2)(|z|^2)^n G_{p,q+1}^{q+1,0}\left(|z|^2 \left| \begin{matrix} / \,; & \{a_i-1\}_1^p \\ 0,\ \{b_j-1\}_1^q \,; & / \end{matrix}\right.\right)=\frac{\prod_{j=1}^{q}\Gamma(b_j)}{\prod_{i=1}^{p}\Gamma(a_i)}n!\frac{\prod_{j=1}^{q}(b_j)_n}{\prod_{i=1}^{p}(a_i)_n} \quad (4.58)$$

let's show that it can also be used to solve some integrals in which an arbitrary function $g\left(x|z|^2\right)$ appears, functions that can be developed in series according to powers of $|z|^2$:

$$g\left(x|z|^2\right)=\sum_{m=0}^{\infty}c_m x^m\left(|z|^2\right)^m \quad (4.59)$$

Consequently, the integral that interests us becomes

***General case 10***

$$\int_0^{R\le\infty} d(|z|^2)\, G_{p,q+1}^{q+1,0}\left(|z|^2 \left| \begin{matrix} / \,; & \{a_i-1\}_1^p \\ 0,\ \{b_j-1\}_1^q \,; & / \end{matrix}\right.\right) g\left(x|z|^2\right)==\sum_{n=0}^{\infty}c_n x^n \times$$

$$\times\int_0^{R\le\infty} d\left(|z|^2\right)\left(|z|^2\right)^n G_{p,q+1}^{q+1,0}\left(|z|^2 \left| \begin{matrix} / \,; & \{a_i-1\}_1^p \\ 0,\ \{b_j-1\}_1^q \,; & / \end{matrix}\right.\right)=\frac{\prod_{j=1}^{q}\Gamma(b_j)}{\prod_{i=1}^{p}\Gamma(a_i)}\sum_{n=0}^{\infty}c_n\, n!\,\frac{\prod_{j=1}^{q}(b_j)_n}{\prod_{i=1}^{p}(a_i)_n}x^n \quad (4.60)$$

The result is considerably simplified if the coefficients $c_n$ are proportional to products of Euler's gamma functions $\Gamma(...)$.

***Example 10.1*** $g\left(x|z|^2\right)=\left(1-x|z|^2\right)^m$

$$\left(1-x|z|^2\right)^m=\sum_{n=0}^{m}\binom{m}{n}(-1)^n x^n\left(|z|^2\right)^n=\sum_{n=0}^{m}(-m)_n\frac{(-x)^n}{n!}\left(|z|^2\right)^n \quad (4.61)$$

$$\int_0^{R\le\infty} d(|z|^2)\, G_{p,q+1}^{q+1,0}\left(|z|^2 \left| \begin{matrix} / \,; & \{a_i-1\}_1^p \\ 0,\ \{b_j-1\}_1^q \,; & / \end{matrix}\right.\right)\left(1-x|z|^2\right)^m=$$

$$=\sum_{n=0}^{m}(-m)_n\frac{(-x)^n}{n!}\int_0^{R\le\infty} d(|z|^2)\left(|z|^2\right)^n G_{p,q+1}^{q+1,0}\left(|z|^2 \left| \begin{matrix} / \,; & \{a_i-1\}_1^p \\ 0,\ \{b_j-1\}_1^q \,; & / \end{matrix}\right.\right)= \quad (4.62)$$

$$=\frac{\prod_{j=1}^{q}\Gamma(b_j)}{\prod_{i=1}^{p}\Gamma(a_i)}\sum_{n=0}^{\infty}(-m)_n\frac{\prod_{j=1}^{q}(b_j)_n}{\prod_{i=1}^{p}(a_i)_n}\frac{(-x)^n}{n!}=\frac{\prod_{j=1}^{q}\Gamma(b_j)}{\prod_{i=1}^{p}\Gamma(a_i)}\,{}_{q+1}F_p\left(-m\,,\,\{b_j\}_1^q\,;\,\{a_i\}_1^p\,;\,x\right)$$

For $a_1 = 1$ and $G_{0,1}^{1,0}\left(|z|^2 \,|\, b-1\right) = \left(|z|^2\right)^{b-1} e^{-|z|^2}$, we obtain

$$\int_0^{R\le\infty} d(|z|^2) G_{0,1}^{1,0}\left(|z|^2 \,|\, b-1\right)\left(1 - x|z|^2\right)^m = \int_0^{R\le\infty} d(|z|^2)\, e^{-|z|^2} \left(|z|^2\right)^{b-1} \left(1 - x|z|^2\right)^m =$$

$$= \sum_{n=0}^{m} (-m)_n \frac{1}{(1)_n} \frac{(-x)^n}{n!} \int_0^{R\le\infty} d(|z|^2) e^{-|z|^2} \left(|z|^2\right)^{n+b-1} = \Gamma(b) \sum_{n=0}^{\infty} (-m)_n \frac{(b)_n}{(1)_n} \frac{(-x)^n}{n!} = \tag{4.63}$$

$$= \Gamma(b)\ {}_2F_1(-m, b; 1; x)$$

while for $b = 1$ we have

$$\int_0^{R\le\infty} d(|z|^2)\, e^{-|z|^2} \left(1 - x|z|^2\right)^m = {}_1F_0(-m\,,\ ;\ ;\ x) = (1 - x\ )^m \tag{4.64}$$

***General case 11.*** Considering the general integral

$$\left(I_{G\times F}\right)_a \quad \int_0^{\infty} d\left(|z|^2\right) G_{p,q+1}^{q+1,0}\left( |z|^2 \left| \begin{matrix} / \,; & \{a_i - 1\}_1^p \\ 0,\ \{b_j - 1\}_1^q\ ; & / \end{matrix} \right.\right) {}_rF_s\left(\{c_i\}_1^r ; \{d_j\}_1^s ; A|z|^2\right) = \tag{4.65}$$
$$= \Gamma_{p,q}(b/a)\ {}_{q+r+1}F_{p+s}\left(\{b_j\}_1^q, \{c_i\}_1^r, 1; \{a_i\}_1^p, \{d_j\}_1^s; A\right)$$

let us calculate the integral in which appear the generalized Laguerre polynomials, with $g\left(|z|^2\right) = {}_1F_1\left(-n\ ; \lambda + 1 ; |z|^2\right) = \dfrac{n!}{(\lambda+1)_n} L_n^{\lambda}\left(|z|^2\right), \quad x = 1$ .

***Example 11.1 -*** $r = 1,\ s = 1,\ c_1 = -n,\ d_1 = \lambda + 1$

$$\int_0^{\infty} d\left(|z|^2\right) G_{p,q+1}^{q+1,0}\left( |z|^2 \left| \begin{matrix} / \,; & \{a_i - 1\}_1^p \\ 0,\ \{b_j - 1\}_1^q\ ; & / \end{matrix} \right.\right) {}_1F_1\left(-n\ ; \lambda + 1 ; |z|^2\right) =$$

$$= \frac{n!}{(\lambda+1)_n} \int_0^{\infty} d\left(|z|^2\right) G_{p,q+1}^{q+1,0}\left( |z|^2 \left| \begin{matrix} / \,; & \{a_i - 1\}_1^p \\ 0,\ \{b_j - 1\}_1^q\ ; & / \end{matrix} \right.\right) L_n^{\lambda}\left(|z|^2\right) = \tag{4.66}$$

$$= \frac{n!}{(\lambda+1)_n} \frac{\prod_{j=1}^{q} \Gamma(b_j)}{\prod_{i=1}^{p} \Gamma(a_i)}\ {}_{q+2}F_{p+1}\left(\{b_j\}_1^q, -n, 1;\ \{a_i\}_1^p, \lambda + 1; 1\right)$$

***Example 11.2 -*** $p = 0,\ q = 0,\ a_1 = 1,\ b_1 = b + 1$

$$G_{0,1}^{1,0}\left(|z|^2 \,|\, b\right) = e^{-|z|^2} \left(|z|^2\right)^b, \quad L_n^{\lambda}\left(|z|^2\right) = \frac{(\lambda+1)_n}{n!} \sum_{m=0}^{n} \frac{(-n)_m}{(\lambda+1)_m} \frac{\left(|z|^2\right)^m}{m!} \tag{4.67}$$

$$\frac{n!}{(\lambda+1)_n}\int_0^\infty d\left(|z|^2\right)e^{-|z|^2}\left(|z|^2\right)^b L_n^\lambda\left(|z|^2\right)=\sum_{m=0}^{n}\frac{(-n)_m}{(\lambda+1)_m}\frac{1}{m!}\int_0^\infty d\left(|z|^2\right)e^{-|z|^2}\left(|z|^2\right)^{b+m}=$$
$$=\Gamma(b+1)\sum_{m=0}^{n}\frac{(b+1)_m(-n)_m}{(\lambda+1)_m}\frac{1}{m!}=\Gamma(b+1)\;{}_2F_1(b+1,-n\;;\;\lambda+1;1) \tag{4.68}$$

and we obtain

$$\int_0^\infty d\left(|z|^2\right)e^{-|z|^2}\left(|z|^2\right)^b L_n^\lambda\left(|z|^2\right)=\frac{\Gamma(b+1)(\lambda+1)_n}{n!}\;{}_2F_1(b+1,-n\;;\;\lambda+1;1) \tag{4.69}$$

which is in agreement with integral 7.414.7, page 809, of the Gradshteyn and Ryshik book [19].

## 5. Concluding remarks

In the paper we show that the coherent states approach will be useful not only in different branches of physics, but also it can be a suitable "feedback" for the theory of special functions. Namely, by using the properties of coherent states, especially equation of the unitary operator decomposition, we lead to a special kind of integrals, which can be called the fundamental coherent states integrals. Beginning from these integrals we have deduced a set of new integrals involving Meijer's and generalized hypergeometric functions. After the particularization of coefficients $p$, $q$, $r$ and $s$, as well as the of set of parameters $\{a_i\}_1^p$, $\{b_j\}_1^q$, $\{c_i\}_1^r$ and $\{d_j\}_1^s$ appearing in these functions, it is possible to recover, on the one hand, a lot of known integrals, and on the other hands to obtain some new unknown integrals. Everywhere we used the *diagonal operators ordering technique* (DOOT), a newly introduced technique related to the normal ordering of the creation and annihilation operators. From the attached examples it is to observe that, from the integration point of view, the creation $\hat{E}_+$, and the annihilation $\hat{E}_-$ operators that appear under the integral's sign can be treated as c-numbers, and consequently they can be replaced by simple numerical constants. The verifications, respectively the verification indications, confirm the correctness of the results that were obtained.